\documentclass[doublecol]{epl2} 
\newcommand{\nc}{\newcommand}
\def\nn{\nonumber\\}
\def\bea{\begin{eqnarray}}
\def\eea{\end{eqnarray}}
    
\nc{\braket}[1]{\langle\,{#1}\rangle}

\def\Z{\mathbf Z}    

\def\V{{\cal V}}

\def\txt{\textstyle}
\def\vperskip{\vspace{5pt}}
\makeatletter\def\epl@stylemark{\hbox to0pt{\vbox to 0pt{\vss \hbox{\sffamily OIQP-06-14} \vskip6ex}}}\makeatother
\title{Effect of interaction shape on the condensed DNA toroid}
\author{Y. Ishimoto\inst{1} \and N. Kikuchi\inst{2,3}}
\institute{
  \inst{1} Okayama Institute for Quantum Physics,
1-9-1 Kyouyama, Okayama 700-0015, Japan\\
  \inst{2} {Institut f\"ur Physik, Johannes Gutenberg-Universit\"at Mainz, Staudinger Weg 7, D-55099 Mainz, Germany}\\
  \inst{3} Fachbereich Physik, Martin-Luther-Universit\"at Halle-Wittenberg, D-06099 Halle, Germany\thanks{
present address.
}
}
\abstract{
We investigate how different microscopic interactions between semiflexible chain segments can qualitatively alter the physical properties of the condensed toroid. We propose a general form of the Hamiltonian of the toroid and discuss its analytic properties. For different interactions, the theory predicts different scaling behaviours of the mean toroidal and cross sectional radii, $r_c$ and $r_{cross}$, as functions of the contour length $L$: $(r_c, r_{cross}) \sim L^{\nu\left(N_c\right)}$ with $\nu=(\frac15, \frac25)$ for the van der Waals type,  $\nu=(-\frac13, \frac23)$ for the Coulomb type,  $\nu=(-1, 1)$ for the delta function type attractions in the asymptotic limit. For the toroids with finite winding number $N_c=100\sim400$, we find $\nu\,\,{\simeq}\,\,0$ for the Yukawa interaction with screening parameter $\kappa=0.5\sim1.0$, and $\nu=0.1\,\,{\sim}\,\,0.13$ for the van der Waals type interactions. These findings could provide possible explanation for the experimentally well known observation $\nu\,\,{\simeq}\,\,0$ of the condensed DNA toroids. Conformational transitions are also discussed.
}

\begin{document}

\maketitle


Condensed DNA toroid and its conformational properties have been of great interest, for instance, as a possible candidate for gene delivery in gene therapy \cite{HDetal,HD01}. Although condensed DNA toroids have been heavily investigated in experiment and theory (\cite{YYK99,SIGPB03,IK06a} and the references therein), it remains unclear what physical factors determine the toroidal sizes \cite{HD01}. In fact, there is experimentally well known observation that the radius of DNA toroid $r_c$ is independent of chain contour length $L=400-50,000 bp$ ($132.8\,nm-16.6\,{\mu}m$): $r_c \sim L^{\nu}$ with $\nu\,\,{\simeq}\,\,0$ \cite{Bloomfield}. In most analytic works, phenomenological free energy models balance the bending and surface free energies to estimate toroidal properties. The predicted scaling relation for the mean toroid radius is $r_c\sim L^{\nu}$ with $\nu=\frac15$ for large $L$ \cite{SIGPB03,L15_lit,MKPW05}. This is however inconsistent to the experimental observation.

In this letter, we show how different microscopic interactions qualitatively modifies physical properties of the toroid, in particular, the mean toroidal and cross sectional radii. 
To achieve this, we first propose a general form of the Hamiltonian of the toroid and discuss its analytic properties. We then specify the explicit forms of attractive interactions, namely delta-function, van der Waals, and Yukawa (screened Coulomb) type interactions. We show exponents of the mean toroidal and cross sectional radii in the large $L$ (or equivalently large dominant winding number $N_c$) limit are categorised into three distinct groups for different interactions. For experimentally realistic winding numbers $N_c=100\sim400$, we find $\nu\,\,{\simeq}\,\,0$ for the Yukawa interaction with inverse screening length $\kappa=0.5\sim1.0$ and $\nu=0.1\,\,{\sim}\,\,0.13$ for the van der Waals type interactions. We also describe conformational transitions.

\section{Theory}
We propose 
the general Hamiltonian of a tightly packed toroid, with any type of short/long-range attractive interactions between segments: 
${\cal H}_{cl}(a,l,L,W) = {\txt \frac{Ll}{2} a^2} - W \Bigl[ \frac{2\pi}{a} \V(N{}) + \left( L - \frac{2\pi N{}}{a} \right) $\\ 
${\txt Gap(N{}) \Bigr]\!}.$
$l$ denotes the persistence length of the semiflexible polymer chain of
total contour length $L$. $l$ is assumed to be large enough relative to the bond length $l_b$ to realise its stiffness ($l \gg l_b$). $a$ is an inverse toroidal radius and $W$ is a positive coupling constant of the two-body attractive interaction between polymer segments. $N{}$ is the winding number: $N{}\equiv [aL/2\pi]$.\footnote{Gauss' symbol $[x]$ gives the greatest integer among $m<x$.}
$\V(N{})$ is the attractive potential in the toroid cross section in the unit of $-W$, which falls into the number of interacting segmental pairs for the van der Waals nearest neighbour interaction \cite{IK06a}: 
$\V(N)=3N-2\sqrt{3}\sqrt{N\!-\!1/4}$.
$Gap(N{})\equiv\V(N{} \!\,+\!1) - \V(N{})$ is introduced to compensate the continuity of the Hamiltonian. 
Temperature $\beta=1/(k_BT)$ is implicitly included in $l$ and $W$. 
An important assumption is that the attractive potential ({\it i.e.} the second term in ${\cal H}_{cl}$) 
is a linear function of the toroid radius, or equivalently of the mean perimeter of the toroid \cite{ALG04}.
With the conformation parameter $c\,\equiv \frac{W}{2l}\!\left(
\frac{L}{2\pi} \right)^{\!2}>0$, and a variable $x = \frac{aL}{2\pi}\geq 0$ ({\it i.e.} $[x] = N{}$), the Hamiltonian is simplified by ${\cal H}_{cl}(a,l,L,W) = WL \cdot {\cal H}(c,x)$ and:
\bea
{\cal H}(c,x) 
 = \frac{x^2}{4c} + \frac{f([x])}{x} -  Gap([x]),
\label{toroidHcx}
\label{Hcx}
\eea
where 
$$
f(N) \equiv N \V(N\!+\!1) - (N\!+\!1) \V(N).  
$$
For consistency, $\V(0) =
\V(1) = 0$, so that $\frac{f([x])}{x} \to 0$ at $x=0$. By varying the
value of $W$, $\V(2)$ can be normalised to unity. For a chain to form a toroid, $WL\gg 1$ and $c>4$ are assumed.

Most physical observables for the tightly packed toroids can be quite accurately estimated by the dominant toroid winding number $N_c$.
For instance, the mean toroid radius is given by 
\bea
r_c\equiv \frac{L}{2\pi N_c}\sim L^{\nu\left(N_c\right)}
\eea
with the exponent $\nu\!\left(N_c\right)$. $N_c$ is given by the global minimum of the Hamiltonian (\ref{Hcx}).
Since ${\cal H}(c,x)$ is governed by the value of $c$, our aim is now to derive the relation between $c$ and $N_c$.

Let us focus on the form of ${\cal H}(c,x)$ in the $m$-th segment ($m < x <m+1$). 
The only extremum is given at $x=x_c(m)\equiv\left( 2c f(m)\right)^{\frac13}$.
The condition for this existence in the segment is $f(m)>0$ and $m < x_c(m) < m+1$, from which it follows 
$c_L(m) < c < c_U(m),$ 
where $c_L(m)=\frac{m^3}{2f(m)}$, $c_U(m)=\frac{(m+1)^3}{2f(m)}$.
Provided $f(m)>0$, 
it means $c_U(m) \simeq c_L(m)$ for large $m$, and therefore, $c \simeq \frac{m^3}{2f(m)}$. Hence, for large $N_c$, the $c$-$N_c$ relation is almost uniquely determined by 
\bea
c \simeq \frac{N_c^3}{2f(N_c)}.
\eea
Note that the second $x$-derivative of ${\cal H}(c,x)$ is positive except at $x\in\Z$ so that all existing extrema are stationary.
It should be emphasised that the $c$-$N_c$ relation for a smaller value of $N_c$ is different from its asymptotic one. This becomes crucial when we compare the theory to experiment in the end: real chains such as DNA have finite length. Instead, we replace the above relation by $c \simeq \frac{c_L(N_c)+c_U(N_c)}{2}$, which works well for the cases examined below. We now provide specific attractive potentials and calculate analytical expressions for the mean toroidal and cross sectional radii.


\vperskip
{\bf The delta function potential:} 
One of the simplest interaction is the delta functional attraction, 
$
{\cal H}_{AT}(s) = -W\!\!\int_0^L\!\!\!ds\!\int_0^{s}\!\!ds^\prime \; \!\!\delta\!\left( \left| \vec{r}(s) - \vec{r}(s^\prime) \right| \right). 
$
In ref.\cite{IK06a}, 
we have shown that the toroid of winding number $N\geq4$ is the stable ground state for $c>4$. 
In this ``ideal toroid'' with zero thickness of the chain, every chain segment interacts equally with all the other segments accumulated on the same arc of the toroid. 
Hence, $\V(N{})$ becomes the number of interacting segmental pairs: 
$$
\V(N)={}_NC_2=\sum_{k=1}^{N-1} k = \frac{1}{2} N(N-1). 
$$
$Gap(N)$ is now given by $Gap(N)=N$. Consequently, $f(N)$ is given by $f(N)=\frac12 N(N+1)>0$. Substituting these into the expressions of $c_{L,U}(N)$, we obtain $c_L(N)=\frac{N^2}{N+1}$ and $c_U(N)=\frac{(N+1)^2}{N}$. In the asymptotic limit, we have 
\bea
N_c \simeq c, 
\eea
as $c_L^{(N_c)}\simeq N_c$ and $c_U^{(N_c)} \simeq N_c$. Thus, the radius of the dominant ideal toroid is 
\bea
r_c = \frac{L}{2\pi N_c} = \frac{4\pi l}{W L}.
\eea

\vperskip
{\bf Van der Waals potential:} In real systems, the toroid has finite thickness hence physics could be different from that of the ideal toroid. Therefore, we must incorporate finite size effect into $\V(N)$ of the Hamiltonian, and then calculate the mean toroidal and cross sectional radii. 

\begin{figure}[t]
\onefigure[width=4.7cm]{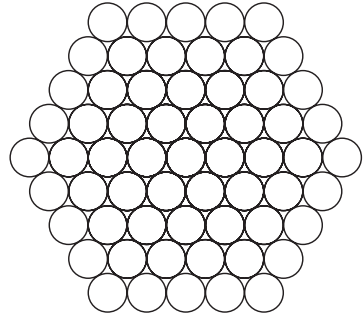}
\caption{A complete hexagon with a side of $5$ segments in the toroidal cross section.}
\label{figHexcross}
\end{figure}

First, we consider the van der Waals type interaction, or equivalently effective short-range dominant attraction. Toroidal cross section can be approximated by the hexagonally arranged chains (Fig.\ref{figHexcross})\cite{IK06a}, interacting via effective nearest neighbour van der Waals attraction. 
By definition, the excluded volume effect of the segments is incorporated in the hexagonally packed cross section. This hexagonal arrangement could be a good approximation as it has been experimentally observed for the condensed DNA toroid \cite{HD01}.
\begin{figure}[t]
\onefigure[width=7.9cm]{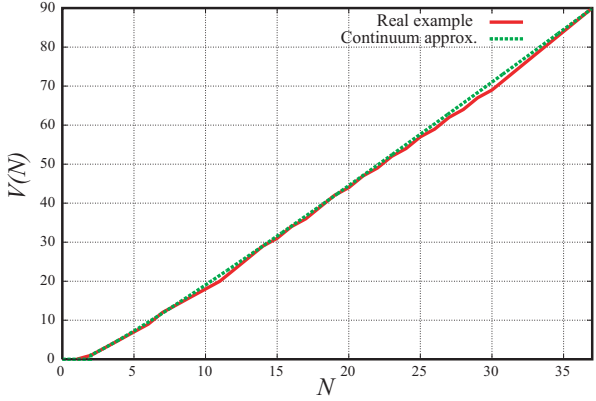}
\caption{An exactly counted discrete function $\V_{discrete}(N)$ and its continuum approximation $\V(N)$ for the number of interacting pairs in the toroidal cross section.}
\label{figHexapprox}
\end{figure}

If the chains are packed in a complete hexagonal cross section (Fig.\ref{figHexcross}),
the winding numbers are $N = 7, 19, 37$, and so on.
In such cases, the number of nearest neighbour interacting pairs between
segments can be counted by the links between neighbouring pairs in the
hexagonal cross section. We then obtain the number as a discrete 
function $\V_{discrete}(N)$. 
In the case of a complete hexagon with a side of $(n+1)$ segments, its winding number is given by 
$$
N=1+\sum_{i=1}^{n} \!6\, i=3n(n+1)+1 , 
$$ 
{\it i.e.}, $n = -\frac12 +\frac{1}{\sqrt{3}}\sqrt{N-\frac14}$. 

When we connect the centres of nearest neighbour segments in Fig.\ref{figHexcross}, it results in a complete hexagon consisting of regular triangular cells with a side $1$. Counting the number of the regular triangles' sides gives the number of the nearest neighbour interaction links: 
\bea&&
\V_{discrete}(N) = \bigl[ 3 \times 
\left({\rm No.~ of~ the~ regular~ triangles\!:}\, 6n^2 \right)
\nn&&\quad
+ \left( {\rm Perimeter~ of~ the~ hexagon\!:}\, 6n \right) 
\bigr]/2
\nn&&
=\left( 3\cdot6n^2+6n\right)/2 =3n(3n+1) 
\nn&&
= 3 N - 2\sqrt{3} \sqrt{N-1/4}, 
\eea
for $N = 1, 7, 19, 37, \cdots$.
Therefore,
for a general value of $N$, we can approximate $\V_{discrete}(N)$ or analytically continue it to the analytic 
function: 
\bea
\V(N) = 3 N - 2\sqrt{3} \sqrt{N-1/4}. 
\eea
This approximates well an exactly counted discrete function $\V_{discrete}(N)$ (see Fig.\ref{figHexapprox}). 
Note that, up to $N=3$ we need not introduce the finite size effect, since there is no difference in the number of interacting pairs between the ideal and the van der Waals nearest neighbour toroids. Thus, $N{}\geq 4$ for this effect.
Note also that the entanglement (knotting) effect of the chain 
arrangement is neglected.

The same analysis presented in the previous section leads to the following ``asymptotic'' $c$-$N_c$ relation of the dominant toroid for large $c$: 
\bea
N_c \simeq \left( 2\sqrt{3} \, c \right)^{\frac{2}{5}}. 
\eea
Substituting this into $r_c\equiv \frac{L}{2\pi N_c}$, we obtain the mean toroid radius: 
\bea
r_c \simeq {\left(6\pi\right)}^{-\frac15}{\left(\frac{l}{W}\right)}^{\!\!\frac25}L^{\frac15}. 
\eea
Note the mean toroid radius of T$4$ DNA in low ionic conditions and Sperm DNA packaged by protamines are quantitatively fitted by this expression \cite{IK06a}. Also, the scaling property $r_{c} \sim L^{\frac15}$ matches the one in \cite{SIGPB03,L15_lit,MKPW05}. 

Similarly, the mean radius of the toroidal cross section can be calculated for the complete hexagonal cross section with a side of $(n+1)$ monomers \cite{IK06a}: 
\bea
r_{cross}
  &\!\!=&\!\! \frac{2+\sqrt{3}}{4}\left(n+\frac12\right)l_d 
\nn
  &\!\!=&\!\! \frac{2\sqrt{3}+3}{12}N^{\frac12}\left[1-\frac{1}{8N} + O\left(\frac{1}{N^2}\right)\right]l_d, 
\eea
where $l_d$ is the diameter of the segment. 
As $N_c \simeq \left( 2\sqrt{3} \, c \right)^{\frac{2}{5}}$ for large $c$, we have 
$$
r_{cross} \simeq \frac{3\sqrt{3}+6}{12}{\left(6\pi\right)}^{\!-\frac25}\!L^{\!\frac25}\! {\left(\frac{W}{l}\right)}^{\!\!\frac15}\!l_d. 
$$
Note that the scaling property $r_{cross} \sim L^{\frac25}$ is in
agreement with the one in \cite{SIGPB03} obtained in the asymptotic limit. Also, we can formally consider the case of the ideal toroid ({\it i.e.} $N_c \simeq c$), although it has zero thickness: $r_{cross} \simeq \frac{2\sqrt{3}+3}{24\pi}L{\left(\frac{W}{2\,l}\right)}^{\!\!\frac12} l_d$.

\section{Yukawa potential and general theory}
In experiment, when we put condensing agents such as multivalent cations into DNA
solution, it can cause DNA to undergo the condensation from a
worm-like chain (whip or coil) to toroidal states \cite{HDetal,HD01,YYK99,SIGPB03,IK06a,Bloomfield,MKPW05}. Due to surrounding
ion clouds, effective interaction between the DNA segments could be described by the screened Coulomb (or Yukawa) potential: 
$$
V_Y(r) = -W\frac{\exp(-\kappa{l_d}
\left(r\!/\!{l_d}\!-1\!\right))}
{\left(r\!/\!{l_d}\right)}, 
$$
where \(r\) is the distance between a pair of segments. 
The interaction range is characterised by the screening
parameter $\kappa$ (inverse screening length), which depends on salt concentrations. In the low screening limit $\kappa \to 0$, the potential corresponds to the Coulomb interaction. 

Not only for the Yukawa potential with various $\kappa$ values, it is in general very difficult to analytically estimate the function $\V(N{})$ just by counting the number of interacting segmental pairs. This is due to the long-range nature of the potential. Thus, we numerically compute the exact value of the attractive potential in the toroidal cross section and do a fit by the following function. We assume that the function $\V(N)$ can be in general expanded as a polynomial in the radius of the cross section ($\propto \left(n+\frac12\right)$): 
\bea
&\V(N) \!\!&\!\!=\! 
  b_0 {\left(\!n\!+\!\frac12\!\right)}^{\!\alpha}\!\!+\! b_1 {\left(\!n\!+\!\frac12\!\right)}^{\!\alpha-1}\!\!+\! b_2 {\left(\!n\!+\!\frac12\!\right)}^{\!\alpha-2}\!\!\nn
  &&\, +b_3{\left(\!n\!+\!\frac12\!\right)}^{\!\alpha-3}\!\!+\! b_4{\left(\!n\!+\!\frac12\!\right)}^{\alpha-4}\!\nn
&\!\!&\!\!=\! A_0{\left(\!N\!-\!\frac14\!\right)}^{\!\!\frac{\alpha}{2}}\!\!\!+\!A_1{\left(\!N\!-\!\frac14\!\right)}^{\!\!\frac{\alpha-1}{2}}\!\!\!\!+\!A_2{\left(\!N\!-\!\frac14\!\right)}^{\!\!\frac{\alpha-2}{2}}\!\!\!\!\nn
&&\, + A_3{\left(\!N\!-\!\frac14\!\right)}^{\!\!\frac{\alpha-3}{2}}\!\!\!\!+\!A_4{\left(\!N\!-\!\frac14\!\right)}^{\!\!\frac{\alpha-4}{2}}\!\!\!,\label{Vn_polynomial}
\eea
where 
$A_0=b_0{\left(\frac{1}{\sqrt3}\right)}^{\!\!\alpha}$, 
$A_1=b_1{\left(\frac{1}{\sqrt3}\right)}^{\!\!\alpha-1}\!\!\!\!\!\!$, 
$A_2=b_2{\left(\frac{1}{\sqrt3}\right)}^{\!\!\alpha-2}\!\!\!\!\!\!$, 
$A_3=b_3{\left(\frac{1}{\sqrt3}\right)}^{\!\!\alpha-3}\!\!\!\!\!\!$, 
$A_4=b_4{\left(\frac{1}{\sqrt3}\right)}^{\!\!\alpha-4}\!\!\!\!\!\!$. 
This fitting is based on the fact that $\V(N)$ of ideal and van der Waals types are given by 
$\V_{ideal}(N) = \frac92{\left(n+\frac12\right)}^{\!4}\!\!-\frac34{\left(n+\frac12\right)}^{2}\!\!-\frac{3}{32}=\frac{N(N-1)}{2}$ 
and 
$\V_{VDW}(N) = 9{\left(n+\frac12\right)}^{2}\!\!-6{\left(n+\frac12\right)}+\frac34=3 N - 2\sqrt{3} \sqrt{N-\frac14}$. 
All segments of the ideal toroid interact with each other with interaction energy $-W$, and $\V(2)$ is normalised to unity. This means the upper bound of $\V(N)$ should be $\V_{ideal}(N)$. On the other hand, the lower bound of $\V(N)$ should be given by the van der Waals nearest neighbour toroid, interacting with energy $-W$. Hence, we have the inequality 
\bea
\V_{ideal}(N)\geq\V(N)\geq\V_{VDW}(N). 
\eea
Therefore, we speculate the Yukawa potential shall be in this region and might have a cubic term in $\left(n+\frac12\right)$ in eq.(\ref{Vn_polynomial}). For the analysis, the cross section has to be small relative to the mean toroid radius. Hence, a toroid-toroidal globule transition point ($r_c \simeq r_{cross} $) may get modified if we consider this effect more seriously. 

The corresponding function $f(N)$ is given by:
\bea
f(N) &\!\!\!=&\!\!\! N \V(N+1) - (N+1) \V(N) 
\nn
&\hspace{-50pt}=&\hspace{-25pt} \sum_iC_i \! \Bigl[ \left(\frac{i}{2}-1\!\right)\!N^{\!\frac{i}{2}}\!+\!\frac{i^2}{16}N^{\!\frac{i}{2}\!-\!1}
\nn&&\hspace{-30pt}
  + \left({\frac{i}{384}\!\left(\frac{i}{2}\!-\!1\right)\left(7i\!-\!34\right)}\!\right)\!N^{\!\frac{i}{2}\!-\!2}
  +\!O\!\left(N^{\!\frac{i}{2}\!-\!3}\right)\Bigr]
\eea
where we sum over $i=\alpha, \alpha\!-1\!, \alpha\!-\!2, \alpha\!-\!3, \alpha\!-\!4$ and 
$C_{\alpha}\!=\!A_0$, $C_{\alpha-1}\!=\!A_1$, $C_{\alpha-2}\!=\!A_2$, $C_{\alpha-3}\!=\!A_3$, $C_{\alpha-4}\!=\!A_4$. 
In the asymptotic limit but below the toroid-toroidal globule transition point, $f(N)$ becomes $A_0\!\left(\frac{\alpha}{2}-1\!\right)\!N^{\!\frac{\alpha}{2}}\!$ for $2< \alpha \leq 4$, 
and $A_1\!\left(\frac{\alpha-1}{2}\!-\!1\!\right)\!N^{\!\frac{\alpha\!-\!1}{2}}\!$ for $\alpha=2$ (VDW type). Note that to satisfy the condition $f(N)>0$ ({\it i.e.} $r_c>0$), we require $C_{\alpha}=A_0\,({\sim}\,b_0)>0$ for $2<\alpha\leq 4$ and $C_{\alpha\!-\!1}=A_1\,({\sim}\,b_1)<0$ for $\alpha=2$.

The mean toroidal ($r_c$) and cross sectional ($r_{cross}$) radii are calculated for large $c$: 
\begin{itemize}
\setlength{\itemsep}{-3pt}
\item[I)] Ideal type ($\alpha=4$): $r_c = \frac{9}{2b_0}\frac{4\pi l}{W\!L}$, $r_{cross}=\frac{2+\sqrt3}{24\pi}\!{\left(\frac{b_0}{3}\right)}^{\!\frac12}\!{\left(\!\frac{W}{l}\!\right)}^{\!\frac12}l_d L$ ($b_0>0$), 
\item[II)] Coulomb type ($\alpha=3$): $r_c = 6\pi^{\!\frac13}b_0^{\!-\!\frac23}\!{\left(\!\frac{l}{W}\!\right)}^{\!\frac23}L^{\!-\!\frac13}$, $r_{cross}=\frac{3+2\sqrt3}{12}{\left(\frac{\sqrt3b_0}{72\pi^2}\right)}^{\!\frac13}\!{\left(\!\frac{W}{l}\!\right)}^{\!\frac13}l_d L^{\!\frac23}$ ($b_0>0$), 
\item[III)] VDW type ($\alpha=2$): $r_c = {\left(\frac{\pi}{6}b_1^2\right)}^{\!-\!\frac15}\!{\left(\!\frac{l}{W}\!\right)}^{\!\frac25}L^{\!\frac15}$, $r_{cross}=\frac{2+\sqrt3}{4}{\left(\frac{-\!b_1}{2^33^3\pi^2}\right)}^{\!\frac15}\!{\left(\!\frac{W}{l}\!\right)}^{\!\frac15}l_dL^{\!\frac25}$ ($b_1<0$). 
\end{itemize} 
Note that the radii of the ideal and the van der Waals nearest neighbour toroids correspond to case I with $b_0=\frac92$ and case III with $b_1=-6$, respectively. We will show below that case II is in fact the Coulomb type.

\def\mh{\hspace{-3pt}}
\begin{table} 
\begin{tabular}{|c|c|c|c|}\hline
$\kappa$ & $\alpha$ & $b_0$ & $b_1$  
\\\hline\hline
0 & 3.00007$\pm$0.00003 & 8.38$\pm$0.00 & -6.25$\pm$0.02 
\\\hline
\mh 0.01 \mh\mh$\!$ & 2.65391$\pm$0.00529& 28.92$\pm$0.74 & -214.27$\pm$13.29 
\\\hline
0.1 & 2.00845$\pm$0.00386 & 110.45$\pm$2.02 & -954.14$\pm$37.64 
\\\hline
0.2 & 2.00048$\pm$0.00207 & 59.20$\pm$0.59 & -259.93$\pm$9.06 
\\\hline
0.3 & 2.00008$\pm$0.00113 & 41.06$\pm$0.22 & -121.61$\pm$3.17 
\\\hline
0.4 & 2.00002$\pm$0.00063 & 32.03$\pm$0.10 & -73.32$\pm$1.34 
\\\hline
0.5 & 1.99957$\pm$0.00020 & 26.69$\pm$0.02 & -51.12$\pm$0.25 
\\\hline
0.6 & 1.99958$\pm$0.00013 & 23.12$\pm$0.01 & -38.56$\pm$0.13 
\\\hline
0.7 & 1.99963$\pm$0.00008 & 20.58$\pm$0.01 & -30.68$\pm$0.08
\\\hline
0.8 & 1.99969$\pm$0.00006 & 18.69$\pm$0.01 & -25.38$\pm$0.05
\\\hline
0.9 & 1.99974$\pm$0.00005 & 17.23$\pm$0.00  & -21.63$\pm$0.04 
\\\hline
1.0 & 1.99978$\pm$0.00004 & 16.07$\pm$0.00 & -18.87$\pm$0.03 
\\\hline
\mh $\beta\!=\!6$ \mh\mh$\!$ & 1.99999$\pm0.00001$ & 9.56$\pm$0.00 & -6.85$\pm$0.00 
\\\hline
\end{tabular}
\caption{Least Squares Fitting of exactly computed $\V(N)$ for the Yukawa potential with various screening parameter $\kappa$ and for $-W{\left(\frac{l_d}r\right)}^{\!\beta}$ with $\beta=6$ using eq.(\ref{Vn_polynomial}). For convenience, $W$ and $l_d$ are taken to be unity. Data is presented for $\alpha$, $b_0$ and $b_1$ only.}
\label{Yukawadata}
\end{table}
Table \ref{Yukawadata} shows the least square fit of exactly computed $\V(N)$ for the potential $-W{\left(\frac{l_d}r\right)}^{\!\beta}$ and the Yukawa potential with inverse screening length $\kappa$ using eq.(\ref{Vn_polynomial}). 
In numerics, we considered the complete hexagon with a side of up to $50$ segments ({\it i.e.} winding number $N=3n(n+1)+1=7351$). The diameter of the chain $l_d$ is taken to be unity for brevity. 
We can confirm that the $\kappa=0$ (Coulomb interaction) result is grouped by case II (Coulomb type). 
The Yukawa potential with $\kappa=0.3, 0.4, 0.5, 0.6, 0.7, 0.8, 0.9, 1.0$ and $\beta=6$ results are classified to case III (VDW type). 
Of great interest is the Yukawa potential with $\kappa=0.01, 0.1, 0.2$, where the exponent $\alpha$ takes the values $2\sim3$. The transition from the Coulomb type (II) to the van der Waals type (III) occurs in this region. We note that for small $n<3$, fit results deviate from $\V(N)$ of the Yukawa with $\kappa\leq 0.3$. However, it does not affect the large $n$ behaviours, thus results reported here sustain. It should be also mentioned that the potential $-W{\left(\frac{l_d}r\right)}^{\!\beta}$ with $\beta=3\sim24$ are categorised by case III (VDW type). But for $\beta=1\sim2$, cubic term becomes important (case II).

\begin{figure}[t]
\hspace{-5pt}
\onefigure[width=8.4cm]{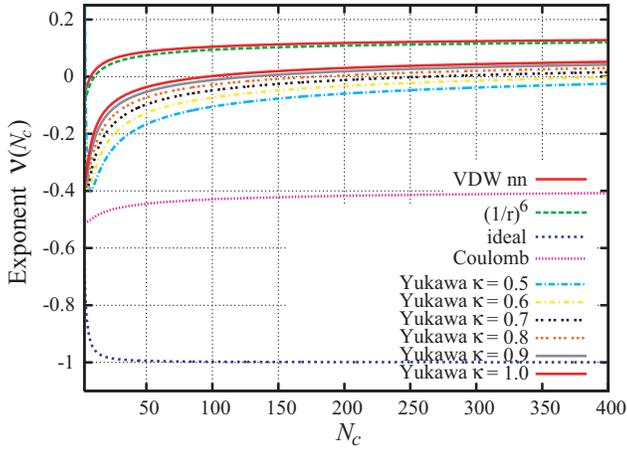}
\vspace{-10pt}
\caption{The exponents $\nu\!\left(N_c\right)$ of the mean toroid radius $r_c\sim L^{\nu\left(N_c\right)}$ for the finite dominant winding number $N_c$ with screening parameter $\kappa=0.5\sim1.0$, Coulomb, full and nearest neighbour Van der Waals, and delta function (ideal) attractions.}
\label{figFinitenu}
\end{figure}

The inequality $\V_{ideal}(N)\geq\V(N)\geq\V_{VDW}(N)$ means the exponents $\nu\!\left(N_c\right)$ for the mean toroid radius $r_c\sim L^{\nu\left(N_c\right)}$ are bounded by those of ideal and van der Waals toroids in the asymptotic limit: 
\bea
-1 \,{\leq}\, \nu\!\left(N_c\right) \,{\leq}\, \frac15. 
\eea
Also, we derive a significant fact that, as far as $\V(N)$ is given as a polynomial in $\left(n+\frac12\right)$, 
we never have $f(N_c){\sim}N_c$ ({\it i.e.} $N_c^2{\sim}c{\sim}L^2$), thus $r_c{\sim}L^0$ for large $c$. This is contradicting to the experimentally well known observation $r_c \sim L^{0}$. To resolve this problem, we plot the exponents $\nu \!\left(N_c\right)$ of the mean toroid radius $r_c\sim L^{\nu \left(N_c\right)}$ for the ``finite'' dominant winding number $N_c$ (Fig.\ref{figFinitenu}). 
The exponents are now defined by 
$\nu\!\left(N_c\right)\!\equiv\!{1-{2}/\!{{\nu_f}(N_c)}}$: 
\begin{eqnarray*}&&
r_c =L/(2\pi{N_c}) 
= L/\!\bigl({2\pi{c^{\frac{1}{{\nu_f}(N_c)}}}}\bigr) 
\nn&&
={{{\left(2\pi\right)}^{\frac{2}{{\nu_f}(N_c)}-1}}}{\left({W}/{2l}\right)}^{\!-\frac{1}{{\nu_f}(N_c)}}\!L^{1-\frac{2}{{\nu_f}(N_c)}}, 
\end{eqnarray*}
where the finite function $\nu_f(N_c)$ is defined as $c= N_c^{\nu_f(N_c)}$. 
We find that for 
$N_c=100\sim400$ ({\it i.e.} realistic winding number of DNA toroids such as T$4$ DNA \cite{YYK99} or Sperm DNA \cite{Bloomfield}), we have $\nu\,\,{\simeq}\,\,0$ for the Yukawa interaction with 
$\kappa=0.5\sim1.0$, and $\nu=0.1\,\,{\sim}\,\,0.13$ for the van der Waals interaction. These could explain the experimental observation: $\nu\,\,{\simeq}\,\,0$.

It is also possible to describe conformational transitions. When the radius of cross section becomes comparable to the segmental diameter $r_{cross} \simeq l_{d}$, the whip-toroid ($N_c=1$) transition occurs. We have $r_c \sim L$ and the transition line $l/W \sim L^2$, which are independent of the interaction shape. Moreover, for the toroid to toroidal globule transition point $r_c \simeq r_{cross} $, we have $r_c \sim l_d^{\frac23}L^{\frac13}$. 
The scaling property $r_c \sim L^{\frac13}$ is similar to the expected scaling of globule like objects. Although the outer radius $r_{outer}$ scales differently for large $c$: $r_{outer}=r_c + r_{cross} \sim r_{cross}$, it scales the same as $r_c$ at the transition point. 

\section{Conclusions}
To summarise, we have shown how different microscopic interactions between chain segments can alter the physical properties of the condensed toroid. In the asymptotic limit, exponents of the mean toroidal and cross sectional radii are categorised into three distinct species: van der Waals type, Coulomb type, and ideal type. For the intermediate winding number of $N_c=100\sim400$, we find $\nu\,\,{\simeq}\,\,0$ for the Yukawa interaction with inverse screening length $\kappa=0.5\sim1.0$, and $\nu=0.1\,\,{\sim}\,\,0.13$ for van der Waals interaction. These finding are consistent to the experimentally well known observation $\nu\,\,{\simeq}\,\,0$. It would be of great interest to check experimentally $r_c \sim L^{\nu}$ for a fixed salt concentration, {\it i.e.}, for a fixed screening parameter $\kappa$. 
If we naively apply the asymptotic values of $r_c$ and $r_{cross}$, transition lines are found to be interaction dependent and to agree with ref. \cite{SIGPB03} in VDW case. However, as has been shown, the asymptotic relations are not so accurate for finite and realistic winding number $N_c$. Therefore, the transition lines are to be studied with a special care.
Finally, It should be stressed that our generic theory can be applied straightforwardly to any toroidal condensation of semiflexible polymer chains with any type and number of microscopic interactions.
\acknowledgments
We are grateful to W. Paul and S. Stepanow for stimulating discussions. We would also like to thank S. Trimper and P. Bruno for stimulating discussions and suggesting investigation of 
$-W{\left(\frac1r\right)}^{\!\beta}$. 
N.K. acknowledges the Deutsche Forschungsgemeinschaft for financial support.

\end{document}